\documentclass[reqno]{JHEP3}
\usepackage{graphicx}
\usepackage{amsmath}
\usepackage{epsfig}
\newcommand{\abs}[1]{\left\vert#1\right\vert}

\newcommand{\zbar}{\overline{z}}
\newcommand{\wbar}{\overline{w}}
\newcommand{\vev}[1]{\langle #1\rangle}
\DeclareMathOperator{\re}{Re} \DeclareMathOperator{\im}{Im}
\DeclareMathOperator{\Tr}{Tr}
\title{Propagators on One-Loop Worldsheets for Supersymmetric Orientifolds and Intersecting Branes}

\author{Friedel T.J. Epple\\Centre for Mathematical Sciences\\
Wilberforce Road\\Cambridge CB3 0WA, UK\\email:
\email{F.Epple@damtp.cam.ac.uk}}

\abstract{We obtain the one-loop propagators in superstring theory
for the general case when the worldsheet fields satisfy
non-trivial holonomy and/or boundary conditions. Non-trivial
holonomy arises in orbifold and orientifold backgrounds whereas
non-trivial boundary conditions arise in backgrounds containing
D-branes of different dimensionality or D-branes intersecting each
other at an angle. In our derivation, we use a generalized version
of the method of images. Dihedral groups play a crucial role in
constructing the one-loop propagators.}

\preprint{DAMTP-2004-116}

\begin{document}

\section{Introduction}

Perturbative string theory is defined in terms of a sum over
worldsheets of different topology. However, the choice of a
particular topology alone does not fix the path integral
completely. Instead, whenever a given worldsheet contains
non-contractible loops, one has to specify the holonomy of the
conformal fields living on the worldsheet. Also, if a worldsheet
contains boundaries, one has to specify boundary conditions.
Depending on the string theory background under consideration,
there can be more than one possible choice for holonomy and
boundary conditions for a given worldsheet. In this case, one has
to sum over all these contributions. Non-trivial holonomy arises
in orbifold and orientifold backgrounds where, if one works in the
covering space, the holonomy group for the bosonic string
coordinates on any given worldsheet is a subgroup of the orbifold
group. Non-standard boundary conditions arise for backgrounds
containing D-branes of different dimensionality or D-branes
intersecting each other at an angle. Taken together, these
configurations are the object of a large part of the existing
literature on string theory backgrounds.

One of the guiding principles in studying string theory
backgrounds has been the search for quasi-realistic backgrounds,
i.e. backgrounds whose low-energy limit is similar to the (Minimal
Supersymmetric) Standard Model or a GUT extension thereof. In this
respect, models with intersecting D-branes or D-branes at orbifold
singularities have been particularly successful (for reviews, see
\cite{Lust:2004ks, Kiritsis:2003mc}). The essential tool for
investigating the low-energy behaviour of a given string theory
background is the low-energy effective action which can be
obtained by computing string scattering amplitudes, i.e. using the
S-matrix approach \cite{Schwarz:1982jn}. At tree-level, the
four-dimensional effective action can be obtained by dimensional
reduction from the ten-dimensional supergravity action. In the
absence of fluxes, it typically contains a large number of moduli.
These can partially be fixed by introducing fluxes on the internal
cycles of a warped Calabi-Yau compactification
\cite{Giddings:2001yu}. Non-perturbative effects have been evoked
to achieve complete moduli stabilization within the framework of
flux compactifications \cite{Kachru:2003aw}. However, it remains
an open question how perturbative corrections change the
picture.\footnote{Reference \cite{Berg:2004ek} studies how
one-loop corrections in the open string sector affects moduli
stabilization, using the background field method}

Computing string scattering amplitudes in the untwisted NS-NS
sector basically requires knowledge of three quantities: the
vertex operators for creating and annihilating asymptotic states,
the vacuum path integral and the propagators of the worldsheet
fields. The vertex operators are completely determined by the
local properties of the underlying CFT and are unaffected by the
choice of topology, holonomy and boundary conditions. On the other
hand, the effect of non-trivial holonomy on the vacuum amplitude
is familiar from the computation of tadpole cancellation
conditions in orientifold backgrounds. There, non-trivial holonomy
is taken into account by including twisted string sectors and/or
inserting twists in the partition function's trace
\cite{Sagnotti:1987tw, Pradisi:1988xd, Ishibashi:1988tf,
Horava:1989vt, Bianchi:1990tb, Bianchi:1990yu, Gimon:1996rq}.
Non-standard boundary conditions affect the partition function
simply by modifying the open string spectrum. It remains to
determine the worldsheet propagators for non-trivial holonomy and
boundary conditions. To fill in this gap is the purpose of this
article.

The method of images is a powerful tool for computing string
theory scattering amplitudes. It has been used to derive
expressions for propagators on tree-level and one-loop worldsheets
with trivial holonomy and boundary conditions
\cite{Burgess:1986ah} and also in direct (tree-level) computations
of tadpoles in type-I string theory \cite{Ohta:1987nq}. In section
\ref{Images}, we review the method of images as applied to the
computation of the standard propagators on one-loop worldsheets
and outline how to generalize this method in order to derive
propagators satisfying generic holonomy and boundary conditions.
In section \ref{propagators}, we apply the generalized method of
images and obtain explicit expressions for the propagators. In
section \ref{discussion}, we discuss applications of our results
and comment on an extension of our methods to the R-R and R-NS
sectors.


\section{The method of images}
\label{Images}

The one-loop surfaces are the torus, the annulus, the Klein bottle
and the M\"obius strip ($\mathcal{T}, \mathcal{A},\mathcal{K}$ and
$\mathcal{M}$). The latter three can be obtained from a covering
torus by identifying points under an antiholomorphic involution
$\mathcal{I}$. This fact has been used to construct the
corresponding propagators with trivial holonomy and NN-boundary
conditions by employing the method of images \cite{Burgess:1986ah,
Antoniadis:1996vw}. We normalize the torus such that
$\mathcal{T}=\mathbb{C}/(\mathbb{Z}+\tau\mathbb{Z})$. Then, table
\ref{construction} summarizes the geometrical construction of
$\mathcal{A}$, $\mathcal{K}$ and $\mathcal{M}$ where we choose the
respective fundamental domain such that the familiar picture of
tubes connecting D-branes and O-planes becomes obvious. In the
case of the Klein bottle, the `loop channel' picture is recovered
by using the fundamental domain $\mathcal{K}=[0,1]\times i[0,t]$
instead, whereas for the M\"obius strip one has to use
$\mathcal{M}=[0,1/2]\times i[0,t/2]$. Here, t denotes the
canonically normalized loop channel proper time (see e.g.
\cite{Angelantonj:2002ct}), i.e. Klein bottle, annulus and
M\"obius strip contribute to the vacuum amplitude through the
terms

\begin{equation}
\mathcal{Z}_{\mathcal{K}}=\int_0^{\infty}
\frac{dt}{2t}\Tr_{cl}\left(\Omega e^{-H_{cl}t}\right);\quad
\mathcal{Z}_{\mathcal{A}}=\int_0^{\infty}
\frac{dt}{2t}\Tr_{o}\left(e^{-H_o t}\right);\quad
\mathcal{Z}_{\mathcal{M}}=\int_0^{\infty}
\frac{dt}{2t}\Tr_{o}\left(\Omega e^{-H_o t}\right)
\end{equation}

respectively where the trace is either over open or closed string
states, $\Omega$ denotes worldsheet orientation reversal and we
implicitly assume implementation of the GSO-projection. The
propagators for trivial holonomy and NN-boundary conditions can be
obtained by symmetrizing the propagator on the covering torus
under the involution $\mathcal{I}$ and restricting to a
fundamental domain of the derived surface. Equivalently, we can
say that for every charge $q$ at a point $z$, one places an
identical mirror charge $q$ at $\mathcal{I}(z)$.

\TABULAR{|c|c|c|c|}{
  \hline
   & $\mathcal{I}(z)$ & $\tau$ & $\mathcal{F}$ \\ \hline
  $\mathcal{A}$ & $-\zbar$ & $it/2$ & $[0,1/2]\times i[0,t/2]$ \\
  $\mathcal{K}$ & $-\zbar+\tau/2$ & $2it$ & $[0,1/2]\times i[0,2t]$ \\
  $\mathcal{M}$ & $-\zbar$ & $it/2+1/2$ & $[1/2,3/4]\times i [0,t]$ \\ \hline
}{The involution $\mathcal{I}$, complex structure $\tau$ of the
covering torus and fundamental domain $\mathcal{F}$ for the
surfaces $\mathcal{A}$, $\mathcal{K}$ and $\mathcal{M}$
respectively. The variable `t' is the canonically normalized loop
modulus \label{construction}}

\EPSFIGURE{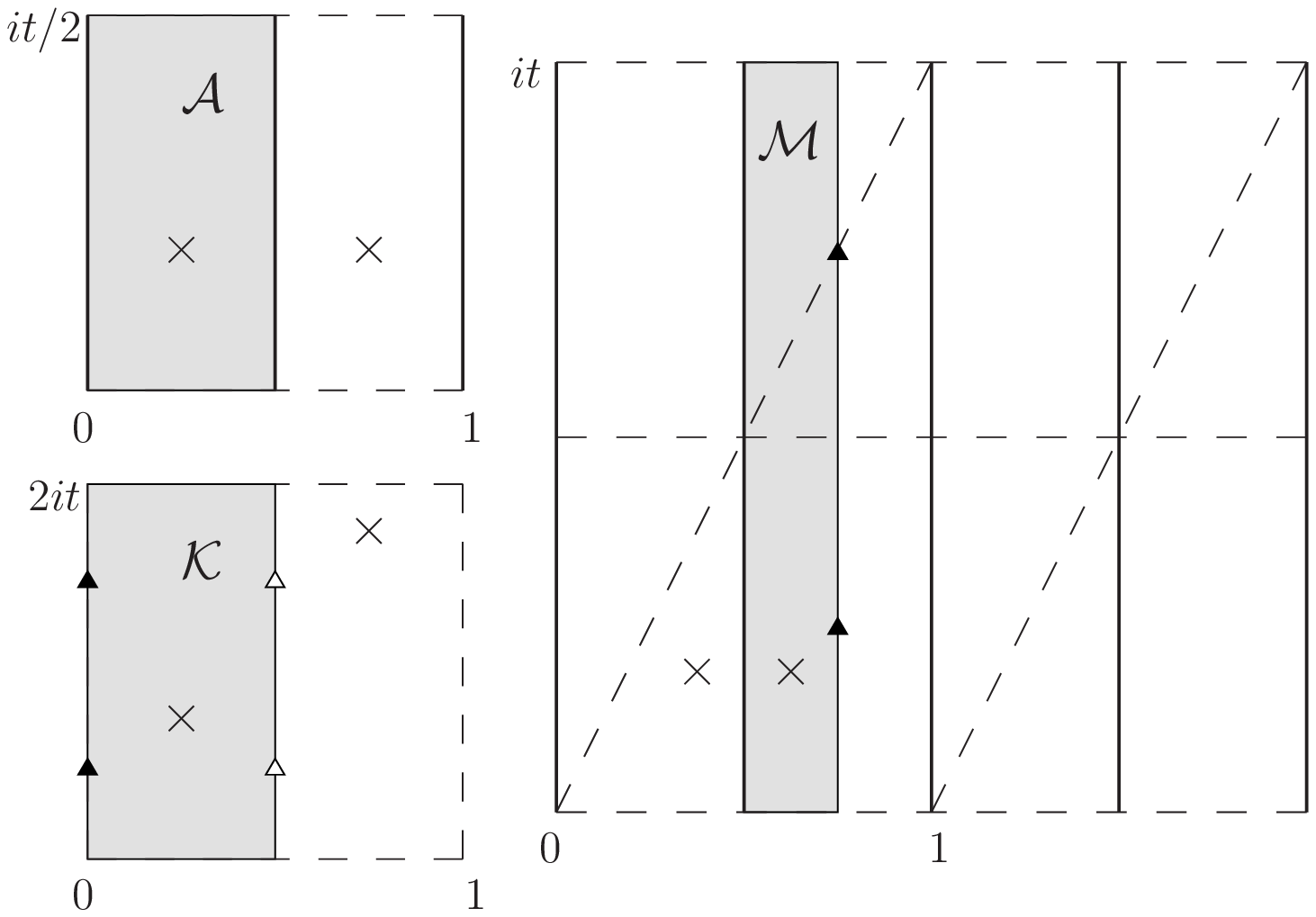}{the annulus, M\"obius strip and Klein bottle
and their respective covering tori. Bold lines mark boundaries,
triangles are to be identified and the x's mark a charge and the
corresponding mirror charge.}


\subsection{Boson propagators}
\label{bosonic}

We generalize the method of images to cover general holonomy as
well as non-trivial boundary conditions. The general strategy is
the following: For a given worldsheet, we pick a minimal set of
closed loops generating the fundamental group. The transformations
of the worldsheet fields when transported along these loops
generate the holonomy group $G$. We differentiate between the
abstract group $G$ and its defining spacetime representation which
we denote by $R$.\footnote{We use the term `representation' in a
loose way. Since we do not require the target space to be a vector
space, strictly speaking we should use the term `left action'
instead.} Then, if we can find a representation $f$ of $G$, acting
on a torus $\mathcal{T}$, we automatically get an induced
representation of $G$ which acts on the worldsheet fields $X$
living on $\mathcal{T}$:

\begin{equation}
g:X^i\mapsto (R_g)^i_{\phantom{i}j} X^j\circ f_g^{-1}.
\end{equation}

Generally, identifying points on the torus under the action of $G$
introduces additional closed loops. Furthermore, $G$-invariant
worldsheet fields pick up a spacetime transformation $R_g$ when
running around the closed loop associated to $g$:

\begin{equation}
X(f_g(z,\zbar))=R_gX(z,\zbar)
\end{equation}

Thus, by symmetrizing the torus propagator under $G$, we can
specifically tailor propagators satisfying a given set of holonomy
conditions when regarded as functions over $\mathcal{T}/G$.
Finally, by choosing an appropriate representation $f$ of $G$, we
ensure that $\mathcal{T}/G$ represents the correct worldsheet.
Generally, we use both holomorphic and antiholomorphic
transformations $f_g$. While holomorphic transformations induce
orientation-preserving loops, antiholomorphic transformations may
induce orientation-reversing loops (i.e. cross-caps) and
boundaries. The latter arise if there are fix points under $f_g$.
In order to deal with worldsheets with boundaries, we have to
further extend the method described so far. Consider a boundary
$b$ and a D-brane $D_b$ which encodes the boundary conditions on
$b$. A crucial observation is the following: If $b$ is given by
the fixed point locus of a transformation $f_b$ acting on a
suitable covering torus, the boundary conditions on $b$ are
equivalent to invariance of the worldsheet fields $X$ under the
transformation

\begin{equation}
\label{boundary} b:X^i\mapsto (R_b)^i_{\phantom{i}j}X^j\circ
f_b^{-1}
\end{equation}

where $R_b$ is a spacetime reflection about $D_b$. This will be
explained in section \ref{annulus} when we discuss the annulus
worldsheet in detail. Thus, by symmetrizing the torus propagator
under \eqref{boundary}, we enforce the correct boundary conditions
at $b$. We are led to the following recipe for obtaining
propagators on a one-loop surface $\sigma$:

\begin{itemize}
\item[(i)] Pick a set of $n$ closed loops which generate the
fundamental group of $\sigma$. Let $g_1,\dots,g_n$ denote the
holonomy transformations along these loops. Also, for each of $m$
boundaries, define a D-brane encoding the boundary conditions and
let $b_1,\dots, b_m$ denote reflections about the respective
D-brane.
\item[(ii)] Define $G=\vev{g_1,\dots,g_n,b_1,\dots,b_m}$. We might
call this the generalized holonomy group as it encodes information
about the holonomy as well as boundary conditions. Find a
representation of $G$ on a suitable torus $\mathcal{T}$ such that
$\sigma=\mathcal{T}/G$.
\item[(iii)] Symmetrize the torus propagator under the induced
action of $G$ and restrict to a fundamental domain of $\sigma$.
\end{itemize}

Finally, note that the second step is not always possible as in
special cases a given worldsheet $\sigma$ cannot be represented as
$\mathcal{T}/G$. As we will see in section \ref{propagators}, this
obstruction does indeed occur if one includes fermionic fields in
the analysis. However, we can solve this problem by a minor
modification of our method. Namely, we can always choose a larger
group $H$, such that $\sigma=\mathcal{T}/H$ and $G=H/N$ where $N$
is a normal subgroup of $H$. In this case, the spacetime
representation of $G$ induces a spacetime representation of $H$
and therefore we get a representation of $H$ on the worldsheet
fields. Consequently, the propagators on $\sigma$ are obtained by
symmetrizing the torus propagator under $H$.


\subsection{Fermion propagators}
\label{fermionprops}

For the fermionic worldsheet fields
$\Psi^{M}(z,\zbar)=(\psi^{M}(z),\tilde{\psi}^{M}(\zbar))$, we have
to allow for a transformation of the spinor indices when going
around closed loops, in addition to any transformation of the
spacetime indices. This means that the holonomy group $\tilde{G}$
for the fermions is generated by transformations

\begin{equation}
\label{fermiontransformation}
{\tilde{g}}:\Psi^i\mapsto
(R_{\tilde{g}})^i_{\phantom{i}j}A_{\tilde{g}}\Psi^j
\end{equation}

where $A$ is a real two-by-two matrix representation of
$\tilde{G}$ acting on the spinor indices and $R:{\tilde{g}}\mapsto
R_{\tilde{g}}$ is a group homomorphism mapping $\tilde{G}$ to the
corresponding holonomy group $G$ for bosonic fields. In analogy to
the bosonic case, we represent $\tilde{G}$ on a suitable covering
torus $\mathcal{T}$ such that $\mathcal{T}/\tilde{G}$ reproduces
the surface of interest. Then, the fermion propagator is obtained
by symmetrizing the torus propagator under the following action of
$\tilde{G}$:

\begin{equation}
{\tilde{g}}:\Psi^i\mapsto (R_{\tilde{g}})^i_{\phantom{i}j}
(A_{\tilde{g}}\Psi)^j\circ f_{\tilde{g}}^{-1},
\end{equation}

where $f$ denotes the representation of $\tilde{G}$ on
$\mathcal{T}$. We will use the Pauli matrices to construct more
explicit expressions for the fermion propagators. For these, we
use the following conventions:

\begin{equation}
\tau_1=\begin{pmatrix}
  0 & +1 \\
  +1 & 0
\end{pmatrix},\qquad
\tau_2=\begin{pmatrix}
  0 & -i \\
  +i & 0
\end{pmatrix},\qquad
\tau_3=\begin{pmatrix}
  +1 & 0 \\
  0 & -1
\end{pmatrix}.
\end{equation}

For a given holonomy group $G$ of the bosonic fields, the group
$\tilde{G}$ is severely constrained by the requirement of
Lorentz-invariance of the RNS action. Going around an
orientation-preserving closed loop, both components of $\Psi$ have
to return to themselves up to the action of some $R_{\tilde{g}}\in
G$ plus a possible sign change which can be assigned to the left-
and right-moving component independently. Therefore, we find that
$A_{\tilde{g}}\in\{\pm\mathbf{1},\pm\tau_3\}$. On the other hand,
going around an orientation-reversing loop relates the left-moving
to the right-moving component. Since going around the loop twice
gives an orientation-preserving loop, the corresponding spinor
transformation has to square to one of the matrices in
$\{\pm\mathbf{1},\pm\tau_3\}$. It follows that, if $\tilde{g}$
corresponds to an orientation-reversing loop, we have
$A_{\tilde{g}}\in\{\pm\tau_1,\pm i\tau_2\}$.\footnote{There can
also be additional factors of $i$ in the matrices associated to
orientation-reversing loops (see e.g. reference
\cite{Blau:1987zb}). We do not make use of this possibility.}.
Finally, boundary conditions can be taken account of by using a
generalized holonomy group in the sense of section \ref{bosonic}.
The transformation of the fermionic fields corresponding to a
boundary $b$ (and a D-brane $D_b$) is restricted by 2-d
supersymmetry and the requirement that boundary terms from varying
the RNS action vanish. If $R_b$ is a reflection about the D-brane
encoding boundary conditions at $b$, the fermionic fields have to
satisfy $\Psi^i=(R_b)^i_{\phantom{i}j}A_b\Psi^j$ on the boundary
where $A_b=\pm\tau_1$. In particular, if we choose coordinates
such that the D-brane is aligned with the coordinate-axes, this
boundary condition reduces to the familiar relation
$\psi=\pm\tilde{\psi}$. Altogether, we find that for $\tilde{g}$
in the \emph{generalized} holonomy group, the fermionic fields
satisfy

\begin{equation}
\Psi^i(f_{\tilde{g}}(z,\zbar))=(R_{\tilde{g}})^i_{\phantom{i}j}A_{\tilde{g}}\Psi^j(z,\zbar)
\label{spinortransformations}
\end{equation}

where the possible choices for $A_{\tilde{g}}$ are summarized as
follows:

\begin{equation}
A\in
\begin{cases}
\{\pm\mathbf{1},\pm\tau_3\}  & \text{ for orientation-preserving loops}\\
\{\pm\tau_1,\pm i\tau_2\} & \text{ for orientation-reversing loops}\\
\{\pm\tau_1\} & \text{ for boundaries}
\end{cases}
\end{equation}

In general, the choice of $A_{\tilde{g}}$ is further constrained
by consistency with the fundamental group (for details, see
section \ref{propagators}). The remaining choices for $A$ account
for the possible spin structures on the respective surface. The
RNS action is invariant under the global symmetry
$\Psi\rightarrow\tau_3\Psi$ which changes the relative sign of
left- and right-moving fermionic modes. In equation
\eqref{spinortransformations}, this symmetry maps $\tau_1$ to
$-\tau_1$ and $\tau_2$ to $-\tau_2$. In the simple examples of
reference \cite{Burgess:1986ah}, only one antiholomorphic
involution per worldsheet was needed whose corresponding sign
could therefore be fixed. In the more general cases considered in
this paper, this is no longer possible as the relative sign
between spinor transformations associated to different
antiholomorphic transformations matters. Only the overall sign can
be fixed.


\subsection{Propagators on covering tori}

We consistently normalize worldsheets such that closed strings in
the `loop channel' have unit length whereas open strings have
length $1/2$. In order to maintain this normalization we need
covering tori of varying size. We use the following notation to
denote tori with non-standard normalization:

\begin{equation}
\label{torusNM} \mathcal{T}^{(N,M)}\equiv\mathbb{C}/
(N\mathbb{Z}+M\tau\mathbb{Z})
\end{equation}

Throughout this paper we will refer to $\tau$ as the complex
structure of $\mathcal{T}^{(N,M)}$. The bosonic propagator on such
a torus is given by

\begin{equation}
\label{propNM}
\vev{X^i(z,\zbar)X^j(w,\overline{w})}_{\mathcal{T}^{N,M}}=
P^{ij}_{N,M}(z,w)=G^{ij}p_{N,M}(z,w)
\end{equation}

where $G^{ij}$ is the spacetime metric and the scalar propagator
$p_{N,M}$ is given by

\begin{equation}
\label{scalarprop}
p_{N,M}(z,w)\equiv -\frac{1}{4}
\ln\abs{\frac{\vartheta_1\left(\frac{z-w}{N}|\frac{M}{N}\tau\right)}{\vartheta'_1\left(0|\frac{M}{N}\tau\right)}}^2
+\frac{\pi}{2NM}\frac{(\im z-\im w)^2}{\im\tau}.
\end{equation}

It can easily be checked that this expression yields the correct
pole structure and periodicity. Similarly, the fermion propagator
for odd spin structure is given by

\begin{equation}
\label{fermionicNM}
\vev{\Psi^i(z,\zbar)(\Psi^j)^{tr}(w,\wbar)}_{\mathcal{T}^{N,M},\text{odd}}
=S^{ij}_{N,M}(z,w)=G^{ij}s_{N,M}(z,w)
\end{equation}

where $\Psi^i(z,\zbar)=(\psi^i(z),\tilde{\psi}^i(\zbar))$ are the
two-dimensional Majorana spinors, the transpose operation refers
to the spinor indices and we define:

\begin{eqnarray}
\label{fermionscalarprop} s_{N,M}\equiv P^F_{N,M}(z,w)
\left(\frac{1+\tau_3}{2}\right)+P^F_{N,M}(\zbar,\wbar)\left(\frac{1-\tau_3}{2}\right)\\
\label{leftfermionprop}
P^F_{N,M}(z,w)\equiv-\frac{1}{4N}\frac{\vartheta_1'\left(\frac{z-w}{N}\Large|\frac{M}{N}\tau\right)}{\vartheta_1\left(\frac{z-w}{N}\Large|\frac{M}{N}\tau\right)}
-\frac{i\pi}{2NM}\frac{(\im z-\im w)}{\im\tau}.
\end{eqnarray}

The bosonic propagator \eqref{propNM} along with the odd spin
structure fermion propagator \eqref{fermionicNM} are sufficient to
express all other propagators in this paper. In particular, we
derive expressions for even spin structure propagators on the
torus using the method of images in appendix \ref{evenSS} and
contrast them to the standard expressions.


\section{Propagators on one-loop surfaces}
\label{propagators}

In this section we derive the propagators for one-loop surfaces
supporting fields with non-trivial holonomy and boundary
conditions. As far as supersymmetric orientifolds are concerned,
our results enjoy full generality, covering Abelian as well as
non-Abelian orientifolds with generic spacetime action. There are
a number of rather special cases, for which the propagators are
already known. Propagators on all untwisted one-loop worldsheets
have been constructed in references \cite{Burgess:1986ah,
Antoniadis:1996vw}. In the appendix of reference
\cite{Bain:2000fb}, the fermion propagator on the twisted torus
for Abelian orbifolds was given in terms of Jacobi functions with
characteristics.\footnote{The boson propagator on twisted tori was
not considered in reference \cite{Bain:2000fb}.} For an
orientifold group of the form $G\times \{1,\Omega\}$, where $G$ is
an Abelian orbifold group, the propagators on some of the
remaining one-loop worldsheets can be obtained from the twisted
torus through a simple application of the method of images. This
was also noted in reference \cite{Bain:2000fb}, where propagators
on the annulus and M\"obius strip were constructed in this way and
used to compute various matter field couplings in the
four-dimensional effective action. However, this approach does not
cover more general orientifold constructions. In particular, it
does not cover the important orientifold constructions widely
known as $\Omega J$-orientifolds and
$\Omega\mathcal{R}$-orientifolds (these arise from orientifold
groups of the form $G\times \mathbb{Z}_2$, where the
$\mathbb{Z}_2$ combines worldsheet orientation reversal with an
inversion of some of the spacetime coordinates). The fermion
propagator on an annulus with DN-boundary conditions was also
given in the appendix of \cite{Bain:2000fb}. In reference
\cite{Antoniadis:2002cs}, the masses of anomalous $U(1)$ gauge
fields in four-dimensional orientifolds were computed, using the
propagators of reference \cite{Bain:2000fb}. We will start by
considering the twisted torus which is the simplest non-trivial
example and serves as an illustration of the generalized method of
images which will be essential in deriving propagators for the
remaining one-loop surfaces.

\subsection{The Torus}

\label{secTorus} We are interested in tori with non-trivial
holonomy.\footnote{Here and elsewhere, we do not mean the holonomy
of the torus itself but rather the holonomy of the corresponding
fibration over the torus of which the worldsheet fields are
sections. We trust that this does not lead to confusion.} These
arise, for instance, in orbifolds of type II string theory on
$\mathbb{R}^{1,3}\times T^6$. Recall that one constructs the
orbifold spectrum in two steps: First, one obtains untwisted
states by projecting the full spectrum onto the subspace of states
which are invariant under some discrete spacetime symmetry
$\mathcal{G}$. Then, one adds the twisted sector which consists of
strings which close onto themselves only up to some symmetry
transformation $g\in \mathcal{G}$. Correspondingly, in the
topological worldsheet expansion, one has to add `twisted tori' on
which the worldsheet fields return to themselves up to the action
of a symmetry group element. Since there are two independent
closed loops on a torus, we can classify all tori by two twists,
$g_1$ and $g_2$. For consistency, these have to be in a
representation of the fundamental group
$\mathbb{Z}\times\mathbb{Z}$ of the torus, i.e. any two twists
appearing in the same worldsheet have to commute with each other.
If we use the familiar parallelogram representation of the torus,
the periodicity conditions for the bosonic worldsheet fields read:

\begin{eqnarray}
X^i(z+1)&=(g_1)^i_{\phantom{i}j} X^j(z)\\
X^i(z+\tau)&=(g_2)^i_{\phantom{i}j} X^j(z)
\end{eqnarray}

For finite $\mathcal{G}$, we have $g_1^N=g_2^M=1$ for some
integers $N$ and $M$ and therefore, the holonomy group $G$ is
given by $\mathbb{Z}_N\times \mathbb{Z}_M$ (or a subgroup
thereof). To obtain the propagator for the twisted torus, we start
with an untwisted torus $\mathcal{T}^{(N,M)}$ [defined in equation
\eqref{torusNM}] on which we represent the action of
$\mathbb{Z}_N\times \mathbb{Z}_M$ by translations along the
lattice directions. Then, using the method of images, we find the
propagator on the twisted torus:

\begin{equation}
\vev{X^i(z,\zbar)X^j(w,\wbar)}_{g_1,g_2}=\sum_{n=0}^{N-1}\sum_{m=0}^{M-1}
(g_1^ng_2^m)^i_{\phantom{i}k} P_{N,M}^{kj}(z-n-m\tau,w)
\label{twistedtorus}
\end{equation}

It is easy to see that this expression satisfies the appropriate
periodicity conditions. Furthermore, it inherits the correct pole
structure from $P_{N,M}$. This procedure works for non-Abelian
orbifold groups only because for a given worldsheet, $g_1$ and
$g_2$ have to commute. Note that because of the fact that the
torus propagator only depends on $z-w$ and moreover is an even
function in $z-w$, we do not need to symmetrize in $w$ separately.

 For the fermionic fields, we have additional freedom in
choosing periodicity conditions. They satisfy

\begin{align}
\Psi^i(z+1)&=(g_1)^i_{\phantom{i}j}A_1\Psi^j(z)\\
\Psi^i(z+\tau)&=(g_2)^i_{\phantom{i}j}A_2\Psi^j(z)
\end{align}

where $A_i\in\{\pm 1,\pm\tau_3\}$ acts on the spinor indices.
Different choices for the matrices $A_i$ represent the 16
different spin structures $(s,\tilde{s})$ on the twisted torus.
Without loss of generality, we can assume $N$ and $M$ to be even
so that the fermionic fields can be analytically continued to
doubly periodic fields on the covering torus
$\mathcal{T}^{(N,M)}$. We should make it clear that this does not
represent a restriction on the form of the orbifold group or its
elements $g_1$ and $g_2$ as $N$ and $M$ are merely unspecified
multiples of the order of $g_1$ and $g_2$ respectively. Since all
$A_i$ mutually commute, we can write the fermion propagator as
follows:

\begin{equation}
\vev{\Psi^i(z,\zbar)(\Psi^{tr})^j(w,\wbar)}_{g_1,g_2,(s,\tilde{s})}
=\sum_{n=0}^{N-1}\sum_{m=0}^{M-1} (g_1^ng_2^m)^i_{\phantom{i}k}
A_1^nA_2^mS_{N,M}^{kj}(z-n-m\tau,w) \label{fermiontwistedtorus}
\end{equation}

In appendix \ref{evenSS}, we show that this expression, when
applied to an untwisted torus, reproduces the familiar even spin
structure fermion propagators. As an application of the general
expressions \eqref{twistedtorus} and \eqref{fermiontwistedtorus}
for the boson and fermion propagators on a twisted torus, we
consider Abelian orbifolds. More precisely, we assume that the six
extra dimensions can be decomposed into pairs of coordinates,
$\mathbf{X}^i\equiv{(X^i,Y^i),\ i=1,2,3}$, such that the orbifold
group acts on these as independent rotations by angles
$\theta^{(i)}$ respectively. Then, the boson propagator takes
block matrix form. Choose integers $N$ and $M$ such that
$N\theta^{(1)}$ and $M\theta^{(2)}$ are multiples of $2\pi$. For a
single pair of coordinates, we obtain the following expression:

\begin{equation}
\vev{\mathbf{X}^{tr}(z,\zbar)\mathbf{X}(w,\wbar)}_{\theta^{(1)},\theta^{(2)}}=
\sum_{n=0}^{N-1}\sum_{m=0}^{M}
R(n\theta^{(1)}+m\theta^{(2)})P_{N,M}(z-n-m\tau,w)\label{abelianTT}
\end{equation}

where $P_{N,M}$ is the boson propagator \eqref{propNM} in the form
of a two-by-two matrix and $R(\theta)$ is the matrix
representation of a rotation by an angle $\theta$. For the
fermions, let $\mathbf{\Psi}^i$ denote the spacetime doublet
corresponding to the bosonic coordinate $\mathbf{X}^i$. Then, the
fermion propagator for a single pair of dimensions takes the form:

\begin{equation}
\vev{\mathbf{\Psi}(z,\zbar)\mathbf{\Psi}^{tr}(w,\wbar)}_{\theta^{(1)},\theta^{(2)},(s,\tilde{s})}
=\sum_{n=0}^{N-1}\sum_{m=0}^{M-1} R(n\theta^{(1)}+m\theta^{(2)})
A_1^nA_2^m S_{N,M}(z-n-m\tau,w) \label{abelianTTfermion}
\end{equation}

where the transpose operation applies to both spinor and spacetime
components and $S_{N,M}$ is the fermion propagator
\eqref{fermionicNM}. Here, we suppress all indices with the
understanding that $R(\theta)$ acts on spacetime indices while the
$A_i$ act on the spinor indices. In appendix \ref{evenSS}, we
compare this expression to the fermion propagator in reference
\cite{Bain:2000fb} and establish equivalence between the two
different expressions.

Finally, we would like to note that for supersymmetry-preserving
orientifolds, we can continue to use expressions \eqref{abelianTT}
and \eqref{abelianTTfermion} after an appropriate coordinate
transformation, even if the orbifold group is non-Abelian. (by
`orbifold group' we mean the subgroup of the orientifold group
consisting of all pure spacetime transformations). All we needed
was that the two group elements $g_1$ and $g_2$ could be written
as independent rotations of the same three coordinate planes.
Equivalently, if we complexify the extra six dimensions, we
require that $g_1$ and $g_2$ are simultaneously diagonalizable by
a unitary coordinate transformation. If the orientifold is to
preserve at least $\mathcal{N}=1$ supersymmetry in four
dimensions, the orbifold group has to be a discrete subgroup of
the R-symmetry $SU(3)$. Each element of $SU(3)$ individually is
unitarily diagonalizable.\footnote{A complex matrix is unitarily
diagonalizable if it commutes with its hermitian conjugate.}
However, as noted above, compatibility with the fundamental group
of the torus means that $g_1$ and $g_2$ have to commute and
therefore they are simultaneously diagonalizable. For non-Abelian
orbifolds, the unitary transformations which diagonalize a given
group element will vary over the full orbifold group and hence the
more general expressions \eqref{twistedtorus} and
\eqref{fermiontwistedtorus} for the propagators on twisted tori
might be more suitable for explicit computations than the
respective specializations \eqref{abelianTT} and
\eqref{abelianTTfermion}. On the other hand, for abelian
orbifolds, one and the same unitary transformation diagonalizes
all group elements.


\subsection{The Klein Bottle}

The Klein Bottle contains two independent loops along which the
worldsheet fields can have non-trivial holonomy. As explained in
section \ref{Images}, the Klein Bottle can be obtained from a
standard torus with complex structure $\tau=2it$ by identifying
points under the antiholomorphic involution
$\mathcal{I}_K:z\mapsto-\zbar+\tau/2$. For our purposes, the
choice $\mathcal{K}=[0,1/2]\times i[0,2t]$ for the fundamental
domain of the Klein Bottle is the most useful as it corresponds to
a tube stretching between two cross-caps. If we consider an
orientifold group of the form $\mathcal{G}+\Omega H$, then going
around a cross-cap in a loop (e.g. going from $0$ to $it$ in our
representation of the Klein bottle), a worldsheet field has to
come back to itself up to some $g\in H$. In the spacetime picture,
we find a corresponding O-plane. On the other hand, if in a given
background, there are O-planes of different type, their
interaction through exchange of closed strings is described by
worldsheets whose cross-cap loops support a holonomy given by two
different group elements, $g_1$ and $g_2$ respectively. These have
to satisfy the consistency condition $g_1^2=g_2^2=h$ for some
$h\in \mathcal{G}$ in order to be compatible with the fundamental
group of the Klein bottle. Following the method of images as
described in section \ref{Images}, let us first determine the
holonomy group $G$ which is generated by $g_1$ and $g_2$. First,
notice that $h$ is in the centre of $G$. We can therefore
construct the quotient group $G_h\equiv G/\vev{h}$. The
equivalence classes $[g_1]$ and $[g_2]$ square to unity and
satisfy $([g_1][g_2])^p=id_{G_h}$ for some integer $p$. These are
the defining relations for the dihedral group $D_p$ (see appendix
\ref{DihedralGroup}), the symmetry group of a regular $p$-sided
polygon and it follows that $G$ is a semidirect product,
$G=\mathbb{Z}_q\rtimes D_p$ where $q$ is the order of $h$. Now, it
is possible to find an appropriate group action of $G$ on the
worldsheet of a rectangular torus $\mathcal{T}^{(p,q)}$ such that
the Klein bottle is given by $\mathcal{T}^{(p,q)}/G$. We represent
the twists $g_1$ and $g_2$ by the following antiholomorphic
worldsheet transformations:

\begin{equation} f_1:z\mapsto -\zbar+\frac{\tau}{2}; \qquad\qquad
f_2:z\mapsto 1-\zbar+\frac{\tau}{2}
\end{equation}

Note that $f_2\circ f_1$ is a translation by $1+\tau$ whereas
$f_1^2=f_2^2$ are simply translations by $\tau$. It is therefore
easy to see that $f_1$ and $f_2$ generate a representation of
$\mathbb{Z}_q\rtimes D_p$. In order to obtain the standard
normalization for the Klein Bottle, we need to set the complex
structure of the covering torus to $\tau=2it$. Then, if we choose
the fundamental domain to be $\mathcal{K}=[0,1/2]\times i[0,2t]$,
the edges $\re z=0$ and $\re z=1/2$ are cross-caps corresponding
to $g_1$ and $g_2$ respectively. The bosonic propagator on the
Klein bottle is given by:

\begin{multline}
\label{KProp} \vev{X^i(z,\zbar),X^j(w,\wbar)}_{g_1,g_2}=
\sum_{n=0}^{p-1}\sum_{m=0}^{q-1}
\Big\{((g_2g_1)^nh^m)^i_{\phantom{i}k} P^{kj}_{p,q}(z+n-\tau (n+m),w) \\
+ (g_1(g_2g_1)^nh^m)^i_{\phantom{i}k} P^{kj}_{p,q}(-\zbar+n-\tau
(n+m+1/2),w)\Big\}
\end{multline}

We define the spin structure for the Klein bottle by

\begin{equation}
\Psi^i(f_1(z))=(g_1)^i_{\phantom{i}j}A_1\Psi^j(z);\qquad
\Psi^i(f_2(z))=(g_2)^i_{\phantom{i}j}A_2\Psi^j(z);
\end{equation}

where the $A_i$ act on the spinor indices. Going around either of
the cross-cap loops twice gives the same orientation-preserving
loop which implies that $A_1^2=A_2^2$ and therefore we must have

\begin{equation}
A_1\in\{\tau_1,i\tau_2\};\qquad A_2= \xi A_1
\end{equation}

with $\xi=\pm 1$. This accounts for all four different spin
structures of the Klein bottle. In order to ensure double
periodicity of the fermion fields on the covering torus, we need
$p$ and $q$ to be even. If this is not the case from the
beginning, we can work with representations of the group
$\mathbb{Z}_{2q} \rtimes D_{2p}$ which contains $\mathbb{Z}_q
\rtimes D_p$ as a subgroup. Hence, without loss of generality, we
may assume that $p$ and $q$ are both even. Then, the fermion
propagator reads:

\begin{multline}
\vev{\Psi^i(z,\zbar),\Psi^j(w,\wbar)}_{g_1,g_2}=
\sum_{n=0}^{p-1}\sum_{m=0}^{q-1}\xi^n
\Big\{((g_2g_1)^nh^m)^i_{\phantom{i}k} A_1^{2n+2m}S^{kj}_{p,q}(z+n-\tau (n+m),w) \\
+ (g_1(g_2g_1)^nh^m)^i_{\phantom{i}k}
A_1^{2n+2m+1}S^{kj}_{p,q}(-\zbar+n-\tau (n+m+1/2),w)\Big\}
\end{multline}

We will now consider the special case of an orientifold which
admits a decomposition of the extra dimensions in pairs of
coordinates $\mathbf{X}^i\equiv(X^i,Y^i)$, such that the
orientifold group acts on these through rotations by angles
$\theta^{(i)}$ respectively. Then, the propagators simplify
considerably. Let us therefore consider a Klein Bottle with
cross-cap twists $g_1$ and $g_2$ which, on a given pair of
worldsheet fields, act as rotations $R(\theta_1)$ and
$R(\theta_2)$ respectively. The Klein Bottle consistency condition
implies $2\theta_1=2\theta_2 \mod 2\pi$ or, equivalently,
$R(\theta_1)=\pm R(\theta_2)$. For the case of
$R(\theta_1)=-R(\theta_2)\equiv R(\theta)$, the scalar propagator
for a pair of worldsheet fields reads:

\begin{align}
\vev{\mathbf{X}(z,\zbar)\mathbf{X}^{tr}(w,\wbar)}_{\theta,-}&=
\sum_{n=0}^{p-1}\sum_{m=0}^{q-1}
(-1)^{n}R(2m\theta)\Big\{P_{p,q}(z-n-\tau m,w)\nonumber\\
&\qquad\qquad+R(\theta)P_{p,q}(-\zbar-n-\tau (m+1/2) ,w)\Big\},
\end{align}

where $P_{p,q}$ is the boson propagator \eqref{propNM} in the form
of a two-by-two matrix and $p$, $q$ are integers such that
$2\theta p =0 \mod 2\pi$ and $(2\theta+\pi) q=0 \mod 2\pi$. If, on
the other hand, $R(\theta_1)=R(\theta_2)\equiv R(\theta)$, the
propagator simplifies even further:

\begin{align}
\vev{\mathbf{X}(z,\zbar)\mathbf{X}^{tr}(w,\wbar)}_{\theta,+}&=
\sum_{m=0}^{q-1} R(2m\theta)\Big\{P_{1,q}(z+\tau m
,w)+R(\theta)P_{1,q}(-\zbar+\tau (m+1/2) ,w)\Big\},
\end{align}

where $q$ is an integer such that $2\theta q=0\mod 2\pi$. The
corresponding fermion propagators are given by (assuming without
loss of generality that $p$ and $q$ are even):

\begin{multline}
\vev{\mathbf{\Psi}(z,\zbar)\mathbf{\Psi}^{tr}(w,\wbar)}_{\theta,\pm}=
\sum_{n=0}^{p-1}\sum_{m=0}^{q-1}
(\pm \xi)^{n}R(2m\theta)A_1^{2n+2m}\Big\{S_{p,q}(z-n-\tau m,w)\\
\qquad\qquad+R(\theta)A_1S_{p,q}(-\zbar-n-\tau (m+1/2) ,w)\Big\}.
\end{multline}

In this expression, $\mathbf{\Psi}$ stands for a pair of fermion
fields, $S_{p,q}$ is the corresponding fermion propagator
\eqref{fermionicNM} with spacetime and spinor indices suppressed,
and it is understood that the rotation matrices $R$ act on
spacetime indices whereas $A_1$ acts on spinor indices.


\subsection{The Annulus}
\label{annulus}

The annulus contains one closed loop and two boundaries. An
obvious modification of the standard annulus propagator is to
accommodate for non-trivial periodicity along the closed loop. We
will come back to the question of twisted annuli later. First,
however, we will focus on different kinds of boundary conditions.
The simplest case is when the fields satisfy the same boundary
conditions, either Dirichlet or Neumann, on both boundaries. As
explained in section \ref{Images}, the annulus can be obtained
from a covering torus by identification under the antiholomorphic
involution $\mathcal{I}:z\mapsto -\zbar$. Boundaries arise as
fixed point loci of $\mathcal{I}$ which are given by $\re z=0$ and
$\re z=1/2$ respectively. One can then use the doubling-trick and
consider periodic worldsheet fields on the covering torus. Neumann
(Dirichlet) boundary conditions are recovered by projecting onto
field configurations which are even (odd) under $\mathcal{I}$. In
terms of mirror charges, this corresponds to placing plus (minus)
one unit of charge at the $\mathcal{I}$-image of the original
charge. There are two common D-brane configurations which require
a generalization of the standard annulus propagators:

\begin{itemize}
\item[(a)] Backgrounds containing D-branes of different
dimensionality
\item[(b)] Backgrounds containing D-branes of different orientations,
intersecting each other at an angle
\end{itemize}

We will start by describing the first case which also serves as a
warm-up for the second, more complicated, case.

\paragraph{D-branes of different dimensionality}

Choosing appropriate coordinates, we can align the D-branes along
the coordinate axes. Besides the familiar NN- and DD-directions,
there will also be ND-directions, for which we have to find the
propagators. Following the mirror charge intuition, we start on a
rectangular torus $\mathcal{T}^{(2,1)}$ with complex structure
$\tau=it/2$ and define the following two involutions:

\begin{equation}
\mathcal{I}_1:z\mapsto -\zbar;\qquad\qquad \mathcal{I}_2:z\mapsto
1-\zbar \label{Z2timesZ2}
\end{equation}

The corresponding fixed point loci are $\mathcal{F}_1=\{z|\re
z=0,1\}$ and $\mathcal{F}_2=\{z|\re z=1/2,3/2\}$ respectively.
Note that $\mathcal{I}_1\circ\mathcal{I}_2$ is a translation by
$1$ and that the two involutions commute up to a lattice shift of
the covering torus. Thus, the two involutions generate the group
$G=\mathbb{Z}_2\times\mathbb{Z}_2$. We can represent the annulus
as $\mathcal{T}^{(2,1)}/G$ and, implying our standard
normalization of the open string length, we choose
$\mathcal{A}=[0,\pi]\times i[0,t/2]$ as the fundamental domain. In
the next step, we introduce positive mirror charges with respect
to $\mathcal{I}_1$ and negative mirror charges with respect to
$\mathcal{I}_2$ to obtain the propagator for ND-boundary
conditions:

\begin{equation}
\vev{X^i(z,\zbar)X^i(w,\wbar)}_{\text{ND}}=
G^{ii}\left\{p_{2,1}(z,w) +p_{2,1}(-\zbar,w) -p_{2,1}(1-\zbar,w)
-p_{2,1}(1+z,w)\right\}.
\end{equation}

where $p$ is the scalar propagator as defined in equation
\eqref{scalarprop}.

\EPSFIGURE{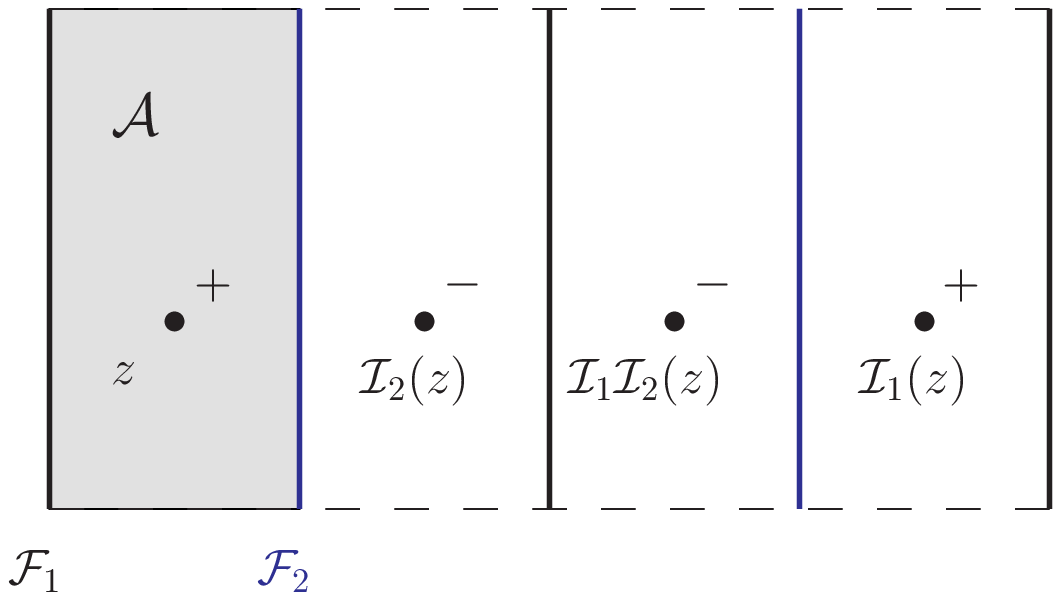}{If the annulus $\mathcal{A}$ is
represented as $\mathcal{T}/(\mathbb{Z}_2\times\mathbb{Z}_2)$, the
bosonic propagator for ND boundary conditions can be obtained by
using three mirror charges of varying signs.}

One can easily check that this propagator satisfies Neumann
conditions on $\mathcal{F}_1$ and Dirichlet conditions on
$\mathcal{F}_2$. For the fermion propagator, we have to consider
different spin structures given by the following relations:

\begin{equation}
\Psi^i(\mathcal{I}_1(z))=A_1\Psi^i(z);\qquad
\Psi^i(\mathcal{I}_2(z))=-A_2\Psi^i(z);\qquad
\Psi^i(\mathcal{J}(z))=A_3\Psi^i(z).
\end{equation}

where $\mathcal{J}$ is the translation $z\rightarrow z+\tau$. As
explained in section \ref{fermionprops}, $A_1$ and $A_2$ must be
chosen from $\pm\tau_1$. Also, their overall sign can be fixed. On
the other hand, $\mathcal{J}$ induces an orientation-preserving
loop, implying $A_3\in\{\pm\mathbf{1},\pm\tau_3\}$. Consistency
requires $A_3$ to commute with both $A_1$ and $A_2$ so that
altogether we find:

\begin{equation}
A_1=\tau_1;\qquad A_2=\pm\tau_1;\qquad A_3=\pm\mathbf{1}.
\end{equation}

The different signs in this equation represent four different spin
structures. We find the fermion propagator by starting with a
torus $\mathcal{T}^{(2,2)}$ and using the method of images:

\begin{multline}
\vev{\Psi^i(z,\zbar)(\Psi^i)^{tr}(w,\wbar)}_{ND}=
G^{ii}\Big\{s_{2,2}(z,w)
+A_1s_{2,2}(-\zbar,w) -A_2s_{2,2}(1-\zbar,w)\\
-A_1A_2s_{2,2}(1+z,w)+ A_3s_{2,2}(\tau+z,w)
+A_1A_3s_{2,2}(\tau-\zbar,w) \\
-A_2A_3s_{2,2}(1+\tau-\zbar,w) -A_1A_2A_3s_{2,2}(1+\tau+z,w)\Big\}
\end{multline}

where $s$ is defined in equation \eqref{fermionscalarprop}. The
fermion propagator on the annulus with ND-boundary conditions was
also given in the appendix of reference \cite{Bain:2000fb} in
terms of Jacobi theta functions with characteristics.

\paragraph{Intersecting D-branes}

Next, we consider two D-branes of the same dimension intersecting
each other. This configuration leads to open string modes which
are localized at the brane intersection. As a consequence,
scattering amplitudes involving an annulus stretching from one
brane to the other renormalize the effective action on the
intersection locus. In the general case, strings stretching
between two intersecting branes obey NN- and DD-boundary
conditions in a number of directions. The remaining dimensions can
be arranged in pairs, $\mathbf{X}^i=(X^i,Y^i)$, on which the two
branes intersect at an angle $\theta_i$ respectively. We choose
coordinates such that the first brane, $D_1$, is aligned along the
$X^i$ axes. Then, the second brane, $D_2$, obeys
$Y_i=\tan(\theta_i) X^i$. For a given pair $\mathbf{X}$ of
worldsheet fields, a non-vanishing angle mixes the correlators of
its components. The propagator for $\mathbf{X}$, written in matrix
notation, acquires off-diagonal elements. Guided by our approach
in the previous paragraphs, we should define two different
antiholomorphic involutions on a covering torus, split the
worldsheet coordinates into parallel and perpendicular components
with respect to the corresponding brane and (anti-)symmetrize the
torus propagator appropriately. More concretely, for a given
worldsheet involution $\mathcal{I}_i$, we define a spacetime
action $R_i$ and combine the two maps into a pair
$g_i\equiv(R_i,\mathcal{I})$ which acts on worldsheet fields as

\begin{equation}
g_i:\mathbf{X}(z)\mapsto (R_i\mathbf{X})(\mathcal{I}_i^{-1}(z))
\end{equation}

If we define the spacetime action to be

\begin{equation}
R_1=\tau_3, \qquad\qquad R_2=R(\theta)\tau_3R(\theta)^{-1},
\end{equation}

where $\tau_3$ is the third Pauli matrix and $R(\theta)$ is a
rotation by $\theta$ in the $(X,Y)$-plane, then
\emph{symmetrizing} a torus propagator under $g_i$ leads to
Dirichlet conditions with respect to $D_i$ on the respective fixed
point locus of $f_i$. However, note that $R_1$ and $R_2$ generally
do not commute and the set $\{\mathbf{1},R_1,R_2\}$ does not close
under matrix multiplication. In fact, $R_1$ and $R_2$ represent
reflections about the respective branes and the product of two
reflections is a rotation,

\begin{equation}
\label{rotref} R_2R_1=R(2\theta).
\end{equation}

Therefore, if the branes intersect at a rational angle, $\theta
\in 2\pi \mathbb{Q}$, the group which is generated by $R_1$ and
$R_2$ is finite. More precisely, if $\theta =\pi p/q$ with two
relatively prime integers, $p<q$, the group $\vev{R_1,R_2}$ is the
dihedral group $D_q$ (see appendix \ref{DihedralGroup}). In order
for $g_1\circ g_2$ to define a group operation, $\mathcal{I}_1$
and $\mathcal{I}_2$ have to be in a representation of $D_q$ as
well. We choose the following representation of $D_q$ on a
covering torus $\mathcal{T}^{(q,1)}$ with complex structure
$\tau=it/2$:

\begin{equation}
\mathcal{I}_1:z\mapsto-\zbar;\qquad\qquad \mathcal{I}_2:z\mapsto
1-\zbar
\end{equation}

Here, $\mathcal{I}_2 \circ \mathcal{I}_1$ is a translation by $1$
and therefore $(\mathcal{I}_2\circ \mathcal{I}_1)^q$ is the
identity on $\mathcal{T}^{(q,1)}$. Finally, we obtain the
propagator for one pair of worldsheet scalars:

\begin{align}
\label{IntersectingProp}
\vev{\mathbf{X}(z,\zbar)\mathbf{X}^{tr}(w,\wbar)}&= \sum_{h\in
D_q}P_{q,1}(h^{-1}z,w)\\
&= \sum_{n=0}^{q-1}R(2\theta n)\left(P_{q,1}(z-n,w)+\tau_3
P_{q,1}(-\zbar+n,w)\right),
\end{align}

where $P_{q,1}$ is the boson propagator \eqref{propNM} in matrix
notation. In order to derive this expression, we generate $D_q$ by
a reflection and a rotation rather than by two different
reflections. A fundamental domain for the annulus is given by
$\mathcal{A}=[0,1/2]\times i[0,t/2]$, which exhibits the correct
open string normalization. As a check, one can verify that our
expression for the propagator satisfies the correct boundary
conditions on $\partial\mathcal{A}$. The procedure for obtaining
the fermion propagators is analogous to the case of D-branes of
different dimensionality. We define the spin structure of the
fermionic fields by

\begin{equation}
\mathbf{\Psi}(\mathcal{I}_1(z))=R_1A_1\mathbf{\Psi}(z);\qquad
\mathbf{\Psi}(\mathcal{I}_2(z))=R_2A_2\mathbf{\Psi}(z);\qquad
\mathbf{\Psi}(\mathcal{J}(z))=A_3\mathbf{\Psi(z)}.
\end{equation}

where $\mathcal{J}$ is the translation $z\rightarrow z+\tau$, the
$R_i$ act on spacetime indices and the $A_i$ act on spinor
indices. Consistency requires:

\begin{equation}
A_1=\tau_1;\qquad A_2=\xi\tau_1;\qquad A_3=\pm\mathbf{1},
\label{intersectingspinstructure}
\end{equation}

with $\xi=\pm 1$. Using equation \eqref{rotref}, it follows that
$\mathbf{\Psi}(z+q)=\xi^q\mathbf{\Psi}(z)$ and therefore, in order
to ensure periodicity on the covering torus, we need $q$ to be
even. If this is not the case from the beginning, we simply drop
the condition that $p$ and $q$ in $\theta=p/q$ be relatively
prime, allowing us to choose $q$ even. Then, the holonomy group
for the bosonic fields will only be a subgroup of $D_q$ but
expression \eqref{IntersectingProp} remains unchanged. The fermion
propagator reads:

\begin{multline}
\vev{\mathbf{\Psi}(z,\zbar)\mathbf{\Psi}^{tr}(w,\wbar)}=
\sum_{k=0}^{q-1}\xi^k\Big\{R(2\theta k)S_{q,2}(z-k,w)+(R(2\theta
k)\tau_3)A_1S_{q,2}(-\zbar+k,w)\\
+R(2\theta k)A_3S_{q,2}(z-k+\tau,w)+(R(2\theta
k)\tau_3)A_3A_1S_{q,2}(-\zbar+k+\tau,w)\Big\}
\label{intersectingfermion}
\end{multline}

where $S_{q,2}$ is the fermion propagator \eqref{fermionicNM} in
matrix form and it is understood that $R(2\theta k)$ acts on
spacetime indices (as well as factors of $\tau_3$ grouped together
with it) whereas the $A_i$ act on spinor indices.

\paragraph{Twisted annuli}

For a twisted annulus, the bosonic fields do not return to
themselves upon going around the closed cycle once but only up to
a transformation $g\in G$ where $G$ is the subgroup of the
orientifold group which contains all pure spacetime
transformations. The path integral on the twisted annulus vanishes
unless both boundaries lie on D-branes which are invariant under
$g$. To see this, consider the loop channel picture where an open
string stretching from one D-brane to another is projected onto
itself after propagating for a time $t$ and being subjected to the
twist.

For any supersymmetry-preserving orientifold, $G$ has to be a
subgroup of $SO(6)$. This means that any given element $g\in G$
can be written as a product of three independent rotations acting
on mutually orthogonal two-dimensional coordinate planes (cf. the
discussion following equation \eqref{abelianTTfermion}). Hence, we
may define pairs of coordinates $\mathbf{X}^i=(X^i,Y^i)$ such that
$g$ acts on these through rotations by an angle $\theta^{(i)}$
respectively. If $G$ is Abelian, we may choose the same coordinate
frame for all possible twists $g$. Otherwise, we have to take
different coordinate frames for different twists. For a given
coordinate pair $\mathbf{X}$, a D-brane can be either
spacetime-filling, point-like or a straight line. The
spacetime-filling and point-like configurations are invariant
under rotations. Combining these leads to boundary conditions of
the NN-, DD- or ND-type. The construction of propagators on
twisted annuli with the respective boundary conditions is fairly
obvious and proceeds along the lines of the twisted torus
construction. Here, we only give the results. For the boson
propagator on a twisted annulus with doubly NN- or DD- boundary
conditions, choose an integer $N$ such that $g$ acts on
$\mathbf{X}$ through a rotation by an angle $\theta=2\pi/N$. Then,
using the method of images, one finds

\begin{equation}
\vev{\mathbf{X}(z,\zbar)\mathbf{X}^{tr}(w,\wbar)}_{\theta,NN/DD}=
\sum_{n=0}^{N-1}R(n\theta)\left(P_{1,N}(z-n\tau,w)\pm
P_{1,N}(-\zbar-n\tau,w)\right)
\end{equation}

Here, $P_{1,N}$ is the boson propagator \eqref{propNM} in matrix
form and $R(\theta)$ is a rotation by an angle $\theta$. The plus
and minus sign apply to NN- and DD-boundary conditions
respectively. The fermion fields can be smoothly continued to
periodic fields on a covering torus $\mathcal{T}^{2,2N}$. The spin
structure on the twisted annulus is defined by

\begin{equation}
\mathbf{\Psi}(-\zbar)=\pm\tau_1\mathbf{\Psi}(z),\qquad
\mathbf{\Psi}(1-\zbar)=\pm \xi_1\tau_1\mathbf{\Psi}(z),\qquad
\mathbf{\Psi}(z+\tau)=\xi_2R(\theta)\mathbf{\Psi}(z)
\end{equation}

where $\mathbf{\Psi}$ is the fermion spacetime doublet
corresponding to the bosonic coordinate $\mathbf{X}$ and it is
understood that $R(\theta)$ acts on spacetime indices. Again, the
plus and minus signs apply to NN- and DD-boundary conditions
respectively. The spin structure is parameterized by the numbers
$\xi_1$ and $\xi_2$ each of which can be either plus or minus one.
Then, the fermion propagator is given by:

\begin{multline}
\vev{\mathbf{\Psi}(z,\zbar)\mathbf{\Psi}^{tr}(w,\wbar)}_{\theta,NN/DD}=
\sum_{n=0}^{2N}\xi_1^nR(n\theta)\Big(AS_{2,2N}(z-n\tau,w)\pm\tau_3S_{2,2N}(-\zbar-n\tau,w)\\
\pm\xi_2\tau_3S_{2,2N}(1-\zbar-n\tau,w)+\xi_2S_{2,2N}(z+1+n\tau,w)\Big)
\label{twistedannulus}
\end{multline}

where $S_{2,2N}$ is the fermion propagator \eqref{fermionicNM} in
matrix form. The propagator \eqref{twistedannulus} can also be
obtained by introducing a single mirror charge on a twisted torus.
This technique was used in reference \cite{Bain:2000fb} to obtain
the fermion propagator in terms of Jacobi theta functions with
characteristics. The boson propagator on the twisted annulus with
doubly ND-boundary conditions is given by

\begin{multline}
\vev{\mathbf{X}(z,\zbar)\mathbf{X}^{tr}(w,\wbar)}_{\theta,ND}=
\sum_{n=0}^{N-1}R(n\theta)\Big(P_{1,N}(z-n\tau,w)+P_{1,N}(-\zbar-n\tau,w)\\
-P_{1,N}(1-\zbar-n\tau,w)-P_{1,N}(z+1-n\tau,w)\Big).
\end{multline}

The spin structure on the twisted ND-annulus is defined by

\begin{equation}
\mathbf{\Psi}(-\zbar)=\tau_1\mathbf{\Psi}(z),\qquad
\mathbf{\Psi}(1-\zbar)=-\xi_1\tau_1\mathbf{\Psi}(z),\qquad
\mathbf{\Psi}(z+\tau)=\xi_2R(\theta)\mathbf{\Psi}(z)
\end{equation}

and the fermion propagator is given by

\begin{multline}
\vev{\mathbf{\Psi}(z,\zbar)\mathbf{\Psi}^{tr}(w,\wbar)}_{\theta,ND}=
\sum_{n=0}^{2N}\xi_1^nR(n\theta)\Big(AS_{2,2N}(z-n\tau,w)+\tau_3S_{2,2N}(-\zbar-n\tau,w)\\
-\xi_2\tau_3S_{2,2N}(1-\zbar-n\tau,w)-\xi_2S_{2,2N}(z+1+n\tau,w)\Big).
\end{multline}

Finally, we consider boundary conditions corresponding to a pair
of D-branes intersecting on the $\mathbf{X}$-plane at an angle
$\phi$. As explained above, the only possible twist in this case
is a rotation by $\pi$, i.e. a $\mathbb{Z}_2$ twist
$g:\mathbf{X}\rightarrow -\mathbf{X}$. Then, if we choose
coordinates such that the first D-brane is aligned with the
X-axis, the boson propagator is given by

\begin{multline}
\vev{\mathbf{X}(z,\zbar)\mathbf{X}^{tr}(w,\wbar)}=
\sum_{n=0}^{q-1}R(2\theta n)\Big(P_{q,2}(z-n,w)+\tau_3
P_{q,2}(-\zbar+n,w)\\
-P_{q,2}(z-n+\tau,w)-\tau_3P_{q,2}(-\zbar+n+\tau,w)\Big),
\end{multline}

where $q$ is an integer such that $q\phi$ is a multiple of $2\pi$.
The spin structure and fermion propagator are given by equations
\eqref{intersectingspinstructure} and \eqref{intersectingfermion}
respectively with the substitution $A_3\rightarrow -A_3$.


\subsection{The M\"obius strip}

The M\"obius strip describes interactions between a D-brane and an
O-plane. The fundamental group of the M\"obius strip is generated
by one cross-cap loop. We choose coordinates such that the twist
associated to the loop can be described as a product of
independent rotations of some coordinate planes
$\mathbf{X}^i\equiv (X^i,Y^i)$ by an angle $\theta_i$
respectively. If for a given coordinate plane, the D-brane we are
concerned with is neither plane-filling nor point-like, we can
rotate the coordinates such that the brane becomes fully aligned
with the X-axis. Thus, we have to distinguish three cases,
corresponding to (N,N)-, (D,D)- and (N,D)-boundary conditions
along the (X,Y)-axes respectively.

\paragraph{(N,N) boundary conditions} For a plane-filling D-brane,
we obtain the propagator by a simple generalization of method
described in references \cite{Burgess:1986ah, Antoniadis:1996vw}.
We start with a skew torus $\mathcal{T}^{(1,q)}$, where
$q\theta=0\mod 2\pi$, with a complex structure parameter
$\tau=1/2+it/2$. Then, identifying points under the involution
$\mathcal{I}:z\mapsto-\zbar$ and the translation
$\mathcal{J}:z\mapsto z+\tau$ introduces a boundary and a
cross-cap. Note that $\mathcal{I}$ and $\mathcal{J}$ commute up to
a $\mathcal{T}^{(1,q)}$-lattice translation. Symmetrizing the
torus propagator under $\mathcal{I}$ leads to Neumann boundary
conditions. On the other hand, symmetrizing under $\mathcal{J}$
with an associated spacetime action corresponding to a rotation by
$\theta$ leads to the correct loop periodicity. Thus, the bosonic
propagator reads:

\begin{equation}
\vev{\mathbf{X}(z,\zbar)\mathbf{X}^{tr}(w,\wbar)}^{(N,N)}_{\theta}=
\sum_{n=0}^{q-1} R(n\theta)\big(P_{1,q}(z-\tau
n,w)+P_{1,q}(-\zbar-\tau n,w)\big)
\end{equation}

We can choose $\mathcal{M}=[0,1/2]\times i[0,t/2]$ as the
fundamental regime for the M\"obius strip. This result can be
regarded as a symmetrization under
$\mathbb{Z}_2\times\mathbb{Z}_q$ where the $\mathbb{Z}_2$-
`reflection' is represented trivially on space-time. The spin
structure of the M\"obius strip is defined by

\begin{equation}
\mathbf{\Psi}(\mathcal{I}z)=A_1\mathbf{\Psi}(z);\qquad
\mathbf{\Psi}(\mathcal{J}z)=R(\theta)A_2\mathbf{\Psi}(z)
\end{equation}

where $R(\theta)$ acts on the spacetime indices and $A_1=\tau_1$
as well as $A_2\in\{\pm\mathbf{1},\pm\tau_3\}$ act on spinor
indices. The fermion propagator reads:

\begin{equation}
\vev{\mathbf{\Psi}(z,\zbar)\mathbf{\Psi}^{tr}(w,\wbar)}^{(N,N)}_{\theta}=
\sum_{n=0}^{2q-1} R(n\theta)A_2^n\big(S_{2,2q}(z-\tau
n,w)+A_1S_{2,2q}(-\zbar-\tau n,w)\big)
\end{equation}

\paragraph{(D,D) boundary conditions} The case of a point-like
brane is very similar. One simply has to anti-symmetrize under
$\mathcal{I}$ to obtain Dirichlet boundary conditions. Note that
in order to retain the correct loop periodicity, one has to assign
a spacetime action $\mathbf{X}\mapsto -R(\theta)\mathbf{X}$ to the
translation $\mathcal{J}$. The bosonic propagator reads:

\begin{equation}
\vev{\mathbf{X}(z,\zbar)\mathbf{X}^{tr}(w,\wbar)}^{D,D}_{\theta}=
\sum_{n=0}^{2q-1} (-1)^nR(n\theta)\big(P_{1,2q}(z-\tau
n,w)-P_{1,2q}(-\zbar-\tau n,w)\big)
\end{equation}

Here, we use a covering torus $\mathcal{T}^{(1,2q)}$ in order to
include the case where $2\pi/\theta$ is odd. For even
$2\pi/\theta$, there is a redundancy in the above expression.
Again, we symmetrize under $\mathbb{Z}_2\times\mathbb{Z}_q$ where
the $\mathbb{Z}_2$ is represented as $\mathbf{X}\mapsto
-\mathbf{X}$ on spacetime. The spin structure is defined by

\begin{equation}
\mathbf{\Psi}(\mathcal{I}z)=A_1\mathbf{\Psi}(z);\qquad
\mathbf{\Psi}(\mathcal{J}z)=-R(\theta)A_2\mathbf{\Psi}(z)
\end{equation}

with $A_1=\tau_1$ and $A_2\in\{\pm\mathbf{1},\pm\tau_3\}$. The
fermion propagator reads:

\begin{equation}
\vev{\mathbf{\Psi}(z,\zbar)\mathbf{\Psi}^{tr}(w,\wbar)}^{(N,N)}_{\theta}=
\sum_{n=0}^{2q-1}(-1)^n R(n\theta)A_2^n\big(S_{2,2q}(z-\tau
n,w)-A_1S_{2,2q}(-\zbar-\tau n,w)\big)
\end{equation}

\paragraph{(N,D) boundary conditions}

Symmetrizing the torus propagator under $\mathcal{I}$ combined
with a space-time reflection about the $X$-axis leads to the
correct boundary conditions. Then, in order to obtain the correct
loop periodicity, one has to combine $\mathcal{J}$ with a rotation
by $\theta$, followed by a reflection about the $X$-axis. This
means that we have to symmetrize under $D_q$. The result is:

\begin{equation}
\vev{\mathbf{X}(z,\zbar)\mathbf{X}^{tr}(w,\wbar)}^{N,D}_{\theta}=
\sum_{n=0}^{2q-1}
\left(\tau_3R(\theta)\right)^n\big(P_{1,2q}(z-\tau
n,w)+\tau_3P_{1,2q}(-\zbar-\tau n,w)\big)
\end{equation}

Using $\tau_3R(\theta)=R(-\theta)\tau_3$, it is easy to check that
the propagator exhibits the correct periodicity. The spin
structure is defined by

\begin{equation}
\mathbf{\Psi}(\mathcal{I}z)=A_1\mathbf{\Psi}(z);\qquad
\mathbf{\Psi}(\mathcal{J}z)=(\tau_3R(\theta))A_2\mathbf{\Psi}(z)
\end{equation}

with $A_1=\tau_1$ and $A_2\in\{\pm\mathbf{1},\pm\tau_3\}$. The
fermion propagator reads:

\begin{equation}
\vev{\mathbf{\Psi}(z,\zbar)\mathbf{\Psi}^{tr}(w,\wbar)}^{(N,N)}_{\theta}=
\sum_{n=0}^{2q-1}(\tau_3^n R(n\theta))A_2^n\big(S_{2,2q}(z-\tau
n,w)+\tau_3A_1S_{2,2q}(-\zbar-\tau n,w)\big)
\end{equation}

In this expression, $\tau_3$ acts on spacetime indices.


\section{Discussion and outlook}
\label{discussion}

In this paper, we have obtained the propagators of the superstring
worldsheet fields on one-loop surfaces for non-trivial holonomy
and boundary conditions. Using our results, it is possible to
compute one-loop corrections to scattering amplitudes in
orientifolds of the type II string theories, including cases where
D-branes of different dimensions or D-branes intersecting at
angles are present. Although our results can be used most directly
in computing scattering amplitudes involving NS-NS states only,
they also apply to interactions with the R-NS and R-R sector. We
will briefly comment on the computation of spin field correlation
functions further down.

From the results for scattering amplitudes, one can obtain the
effective action for the background in question using the S-matrix
approach. Orientifolds with coincident stacks of D-branes and
orientifolds with intersecting D-branes constitute a large class
of perturbatively consistent string theory vacua. Furthermore, a
great number of phenomenologically attractive orientifold models
have been constructed which come quite close to having realistic
low-energy behaviour. Orientifold models are seriously constrained
by consistency conditions imposed by tadpole cancellation. The
only known models with full tadpole cancellation are
supersymmetric orientifolds. While the superpotential is protected
from perturbative corrections by supersymmetry, the K\"ahler
potential does receive corrections at one-loop. The latter might
be interesting in terms of moduli stabilization. For instance,
one-loop corrections might lead to a non-trivial dependence of the
K\"ahler potential on the overall size of the extra dimensions and
to a non-trivial scalar potential for the corresponding K\"ahler
modulus. One-loop corrections might also help to stabilize the
string coupling at small values. This prospect would offer a
detailed view on the string theory landscape in a particular
exactly calculable region.

As a step in this direction, we propose to expand on the
computation of the induced Einstein-Hilbert term for intersecting
D-branes \cite{Epple:2004ra} or D-branes at orbifold singularities
\cite{Kohlprath:2003pu} (see also references
\cite{Kiritsis:2001bc, Antoniadis:1997eg, Antoniadis:2002tr,
Antoniadis:2003sw}). The respective computations involve
two-graviton scattering on one-loop surfaces with the gravitons
polarized along the non-compact directions. The computation of
metric moduli interaction terms would be similar in form but with
polarizations transverse to the non-compact directions and thereby
requiring propagators of the form we derived in this paper.

A complete characterization of the low-energy effective action
requires knowledge of R-R and R-NS scattering amplitudes. Here, we
briefly point out how these can be computed using the methods
developed in this paper. Vertex operators in Ramond sectors are
constructed using spin fields \cite{Friedan:1985ge}.
Unfortunately, this means that it is no longer possible to use
Wick contractions in order to directly reduce the computation of
general scattering amplitudes to the computation of free field
propagators. Instead, one needs the full n-point functions
involving an arbitrary number of worldsheet fermions and spin
fields. These can also be obtained using the method of images as
developed in this paper. The spin fields are given by
$S(z)=C_{\lambda} e^{i\lambda_i\phi_i}$ where $\lambda$ is a
polarization spinor and $\phi_i$ are the bosonized worldsheet
fermions, $e^{i\phi_i}\cong 2^{-1/2}(\psi^{2i}+i\psi^{2i+1})$. $C$
is a cocycle which imposes the correct anticommutation relations.
The holonomy and boundary conditions for the fermion fields
$\psi^{\mu}$ and $\tilde{\psi}^{\mu}$ can be translated to
conditions for their bosonizations $\phi_i$ and $\tilde{\phi}_i$.
These can be rewritten as invariance of the boson doublet
$\Phi_i\equiv(\phi_i,\tilde{\phi}_i)$ under the (generalized)
holonomy group $\tilde{G}$, i.e. $\tilde{g}\Phi=\Phi$ where the
action of $\tilde{g}$ is determined by the corresponding fermion
transformation \eqref{fermiontransformation}. Although in general
the action of $\tilde{G}$ on $\Phi$ will be given by highly
non-trivial relations, it can easily be obtained in the special
case where the spacetime action of $\tilde{G}$ independently
rotates pairs of coordinates: spacetime rotations are replaced by
shifts of the bosonized fields. Having derived the action of
$\tilde{G}$ on $\Phi$, one can proceed along the lines of section
\ref{propagators} and compute any fermion n-point function on a
one-loop surface $\sigma$ by reconstructing the worldsheet from a
covering torus as $\sigma=\mathcal{T}/\tilde{G}$ and symmetrizing
the propagators for the bosonized fields under the action of
$\tilde{G}$. Using this method, the spin field correlators can be
derived from the correlators on the untwisted torus which can be
found in reference \cite{Atick:1986ns}.

A complete characterization of the low-energy effective action
would also involve computing scattering amplitudes with twisted
closed strings and open strings either obeying ND-boundary
conditions or stretching from one brane to another brane which
intersects the first brane at an angle. In all three cases, vertex
operators have been constructed using twist fields. Scattering
amplitudes involving twist fields at tree-level have been computed
for twisted closed strings \cite{Dixon:1986qv}, for open
ND-strings \cite{Frohlich:1999ss} and intersecting branes
\cite{Cvetic:2003ch, Abel:2003vv, Abel:2003yx}. The corresponding
methods would have to be appropriately extended and generalized.


\acknowledgments I am grateful to Mattias Wohlfarth for a
stimulating discussion in the early stages of the project. I would
also like to thank Aninda Sinha and Michael Green for their
support. This work was funded by EPSRC and a Gates Cambridge
Scholarship.


\appendix


\section{Dihedral Groups}
\label{DihedralGroup}

The $q$-th Dihedral Group, $D_q$, is a non-Abelian permutation
group of order 2q. It is the symmetry group of a regular q-sided
polygon, consisting of reflections and rotations. The $q$-th
Dihedral Group can be defined abstractly by

\FIGURE[htb]{
\includegraphics{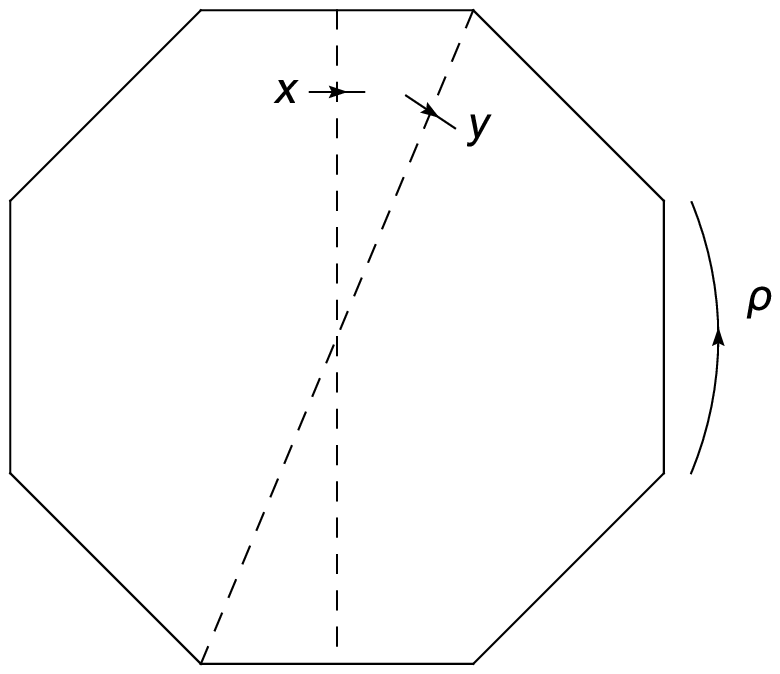}
\caption{The polyhedral group $D_8$. It can be generated by two
reflections, $x$ and $y$, or by one reflection and a rotation,
$\rho$.} }

\begin{equation}
D_q\equiv\vev{x,y|x^2=y^2=(xy)^q=1}.
\end{equation}

In the geometrical picture, $x$ and $y$ can be taken to be
reflections about neighbouring symmetry axes of the polygon.
Alternatively, the Dihedral group can be presented in the
following way:

\begin{equation}
D_q\equiv\vev{x,\rho|x^2=\rho^q=1;x\rho x^{-1}=\rho^{-1}}.
\end{equation}

The latter version is related to the first one by setting
$\rho=xy$. Geometrically, this means that the symmetry group of a
regular q-sided polygon can also be generated by a reflection and
a rotation. The subgroup $\mathbb{Z}_q=\vev{\rho}$ is normal in
$D_q$ which can therefore be written as a semidirect product:

\begin{equation}
D_q\equiv \mathbb{Z}_q \rtimes \mathbb{Z}_2
\end{equation}

In this paper we basically use two different ways for $D_q$ to
operate on a torus. The simplest group action is the one where
$D_q$ operates naturally on one of the circles (picture a polygon
inscribed in the circle) and leaves the other circle invariant.
This generically implies two different fixed point loci
corresponding to reflections about symmetry axes which cut the
polygon at its corners and the middle of its sides respectively.
Thus, we achieve different boundary conditions for the two
boundaries of an annulus. For special values of the torus' complex
structure, both boundaries are joined and we obtain a
representation of the M\"obius strip. Another useful
representation of the group action is obtained by assigning an
additional shift along half of the second circle to all
reflections. This is compatible with the group structure and as a
consequence of this modification, there are no fix-points under
the group action. Therefore, the latter group action is useful in
constructing propagators on the M\"obius strip and Klein bottle
worldsheet.


\section{Comparison to existing results.}
\label{evenSS}

Using the method of images, one can easily obtain expressions for
the fermion propagators on an untwisted torus with non-trivial
spin structure s (this was first suggested in
\cite{Burgess:1986ah}). For the left-moving mode, the propagators
are given by

\begin{equation}
\label{prop1}
\vev{\psi(z)\psi(w)}_s=P^F_{2,2}(z,w)+a_sP^F_{2,2}(z+1,w)
+b_sP^F_{2,2}(z+\tau,w)+a_sb_sP^F_{2,2}(z+1+\tau,w)
\end{equation}

\TABULAR{|c||c|c||c|c|}{
 \hline
 s & a & b & $\alpha$ & $\beta$\\
 \hline
 1 & 1 & 1 & 0 & 0\\
 2 & 1 & -1 & 0 & $\frac{1}{2}$\\
 3 & -1 & -1 & $\frac{1}{2}$ & $\frac{1}{2}$\\
 4 & -1 & 1 & $\frac{1}{2}$ & 0\\
 \hline
}{\label{ssperiods} spin structure conventions }

where $P^F$ is the fermion propagator \eqref{leftfermionprop} and
the conventional relation of $(a,b)$ to the spin structure s is
given in table \ref{ssperiods}. By construction, this expression
has the right periodicity and pole structure if one constrains it
to a fundamental domain of the torus $\mathcal{T}^{(1,1)}$.
However, on first sight it seems to be different from the more
familiar expression (see e.g. \cite{Antoniadis:1996vw})

\begin{equation}
\label{prop2}
\vev{\psi(z)\psi(w)}_s=-\frac{1}{4}\frac{\vartheta_s(z-w|\tau)\vartheta_1'(0|\tau)}{\vartheta_s(0|\tau)\vartheta_1(z-w|\tau)}
\end{equation}

for fermion propagators on a torus with even spin structure
($s=2,3,4$). That the two different expressions are actually equal
follows from the following argument: The propagators \eqref{prop1}
and \eqref{prop2} are elliptic functions (i.e. meromorphic and
doubly periodic) with periods $2$ and $2\tau$. Both propagators
have four simple poles per fundamental cell. In both cases, the
poles are at $z-w=0,1,\tau,1+\tau$ with residues $-1/4, -a_s/4,
-b_s/4, -a_sb_s/4$ respectively, i.e. the infinite part of both
expressions is identical. It follows that the difference of the
two propagators is an analytic function on the whole complex plain
which is bounded and therefore constant by Liouville's theorem. By
evaluating the two expressions at a particular point, one
establishes that this constant vanishes and that the two
expressions are equal.

A similar argument also holds for the twisted torus in Abelian
orbifolds. Consider a pair of coordinates $\mathbf{X}=(X_1,X_2)$
and $\mathbf{\Psi}=(\Psi_1,\Psi_2)$ such that a given element
$g_i$ of the orbifold group acts on these through rotations by an
angle $\theta_1$. For the twisted torus, there are two independent
closed loops along which the worldsheet fields return to
themselves up to the action of group elements $\theta_2$ and
$\theta_2$ respectively. Let us consider the fermionic fields
first. For these, the periodicity conditions may include an
additional sign change. Therefore, we are interested in fermionic
fields $\mathbf{\Psi}(z)$ on the torus
$\mathcal{T}=\mathbb{C}/(\mathbb{Z}\times\tau\mathbb{Z})$ with
periodicity conditions

\begin{equation}
\mathbf{\Psi}(z+1)=A_1R(\theta_1)\mathbf{\Psi}(z), \qquad
\mathbf{\Psi}(z+\tau)=A_2R(\theta_2)\mathbf{\Psi}(z),
\label{TTspinstructure}
\end{equation}

Here, the $A_i\in\{1,\tau_3\}$ act on the spinor indices of
$\mathbf{\Psi}$ whereas $R(\theta)$ denotes a spacetime rotation
by an angle $\theta$. Then, as discussed in section
\ref{secTorus}, if we choose even integers $N$ and $M$ such that
$N\theta_1$ and $M\theta_2$ are integer multiples of $2\pi$, the
fermion propagator takes the following form:

\begin{equation}
\vev{\mathbf{\Psi}(z,\zbar)\mathbf{\Psi}^{tr}(w,\wbar)}_{\theta_1,\theta_2,(s,\tilde{s})}
=\sum_{n=0}^{N-1}\sum_{m=0}^{M-1} R(n\theta_1+m\theta_2)
A_1^nA_2^m S_{N,M}(z-n-m\tau,w), \label{abelianTTfermionapp}
\end{equation}

where $(s,\tilde{s})$ denotes the choice of spin structure
\eqref{TTspinstructure} and $S_{N,M}$ is the fermion propagator
\eqref{fermionicNM}. In order to compare this expression to the
fermion propagator in reference \cite{Bain:2000fb}, we go to the
component notation for the fermionic fields. Let

\begin{equation}
\psi(z)=\frac{1}{\sqrt{2}}\left(\psi^1(z)+i\psi^2(z)\right),\qquad
\overline{\psi}(z)=\frac{1}{\sqrt{2}}\left(\psi^1(z)-i\psi^2(z)\right)
\end{equation}

be the complexified left-handed part of the fermionic coordinates.
Furthermore, define $a=(A_1)_{11}$ and $b=(A_2)_{11}$. In terms of
these, the periodicity conditions \eqref{TTspinstructure} for the
left-handed fermion fields take the following form:

\begin{align}
\psi(z+1)&=ae^{i\theta_1}\psi(z),\qquad
\overline{\psi}(z+1)=ae^{-i\theta_1}\overline{\psi}(z)\\
\psi(z+\tau)&=be^{i\theta_2}\psi(z),\qquad
\overline{\psi}(z+\tau)=be^{-i\theta_2}\overline{\psi}(z)
\end{align}

and likewise for the right-handed components. Then, if the pair of
directions $(X_1,X_2)$ form a square torus, i.e. if for
$i,j\in\{1,2\}$ we have a target space metric $G_{ij}=\eta_{ij}$
(after an appropriate renormalization), the fermion propagators
\eqref{abelianTTfermion} take the following form in complex
notation:

\begin{equation}
\vev{\psi(z)\overline{\psi}(w)}_{\theta_1,\theta_2,s}
=\sum_{n=0}^{N-1}\sum_{m=0}^{M-1}
e^{in\theta_1+im\theta_2}a^nb^mP^F_{N,M}(z-n-m\tau,w)
\label{TTimages}
\end{equation}

All other correlators vanish. By construction, the fermion
propagator, if viewed as a function on the unit torus, has a
single pole at $z=w$ with a residue of $-\frac{1}{4}$.
Furthermore, it picks up a phase $ae^{i\theta_1}$ under
$z\rightarrow z+1$ and a phase $be^{i\theta_1}$ under
$z\rightarrow z+\tau$. The same is true for the expression

\begin{equation}
-\frac{1}{4}\frac{\vartheta[^{\alpha+v_1}_{\beta+v_2}](z-w|\tau)\vartheta_1'(0|\tau)}
{\vartheta[^{\alpha+v_1}_{\beta+v_2}](0|\tau)\vartheta_1(z-w|\tau)}
\label{TTcharacteristics}
\end{equation}

involving Jacobi's theta functions with characteristics. Here,
$\alpha$ and $\beta$ can be read off table \ref{ssperiods} and we
define $v_i\equiv\theta_i/2\pi$. The same expression, in a
different normalization, was used in reference \cite{Bain:2000fb},
with $v_1=0$. Equality between expressions \eqref{TTimages} and
\eqref{TTcharacteristics} can be established using an argument
equivalent to the one given above for fermion propagators on an
untwisted torus with even spin-structure. Starting with the
expression \eqref{TTcharacteristics} with $v_1=0$, it is
straightforward to derive expressions for the twisted annulus and
M\"obius strip using the simple method of images as described in
the appendix of reference \cite{Antoniadis:1996vw}. The resulting
expressions were used to compute various matter field couplings
\cite{Bain:2000fb} and the masses of anomalous $U(1)$ gauge bosons
\cite{Antoniadis:2002cs} in four-dimensional orientifold vacua. It
is also possible to obtain, in this way, the propagators for a
Klein bottle which is twisted only in one direction. However, the
same approach is unable to produce results for the most general
Klein bottle, i.e. the Klein bottle with two independent twists.


\bibliographystyle{JHEP-2}

\providecommand{\href}[2]{#2}\begingroup\raggedright\endgroup

\end{document}